\documentstyle[preprint,prl,aps]{revtex}
\newcommand{\pr}{{\it Phys. Rev.}}

\begin{document}
\draft

\title{
The Hubbard model with smooth boundary conditions
}
\author{ M.\ Veki\'c\cite{deceased} and S.R.\ White }
\address{
Department of Physics,
University of California
Irvine, CA 92717
} 
\date{\today}
\maketitle
\begin{abstract}
We apply recently developed smooth boundary conditions to the 
quantum Monte Carlo simulation of the two-dimensional Hubbard model.
At half-filling, where there is no sign problem, we show that the
thermodynamic limit is reached more rapidly with smooth rather than with
periodic or open boundary conditions. 
Away from half-filling, where ordinarily the
simulation cannot be carried out at low temperatures due to the existence of
the sign problem, we show that smooth boundary conditions allow us to
reach significantly lower temperatures. We examine pairing correlation 
functions away from half-filling
in order to determine the possible existence of a superconducting
state. 
On a $10\times 10$ lattice 
for $U=4$, at a filling of $\langle n \rangle = 0.87$ and
an inverse temperature of
$\beta=10$, we did find enhancement of the $d$-wave
correlations with respect to the non-interacting case, a possible
sign of $d$-wave superconductivity.
\end{abstract}


\narrowtext

\section{Introduction}

The two-dimensional (2D) Hubbard model \cite{Hubbard} is considered to
be one of the possible models to describe the high $T_c$ copper-oxide
superconductors \cite{highTc}. Despite the fact that this model
describes quite well the insulating state and some of the normal state
properties of the copper-oxides, it is not clear whether it contains all
the necessary microscopic features to also explain their superconducting
properties \cite{AF}.  Although a variety of approximate calculations
\cite{Anderson,Hirsch,Loh} predict the existence of superconductivity in
the doped Hubbard model, there is still need to find confirmation from a
calculation which does not depend on uncontrolled approximations. 
 
Quantum Monte Carlo (QMC) simulations have emerged as a method of
choice in trying to solve this model numerically, since, in principle,
the interaction is treated in an exact way \cite{HubMC}.  The path
integral formalism is the standard starting point, where the partition
function is decomposed by using the Suzuki-Trotter formula
\cite{Suzuki}.  A widely used algorithm employs the Hubbard-Stratonovich
transformation to decouple the interaction and integrate out the
fermionic fields \cite{HubStrat}, leading to a bosonic path integral
with an effective action which depends on the fermion determinant given
in terms of the bosonic fields only. The trace over the Stratonovich
fields can then be replaced by a stochastic sampling in which the
fermion determinant becomes essentially the weight of the distibution of
Hubbard-Stratonovich fields \cite{boseaction}.

However, these simulations have a number of hardware limitations that
cannot be easily overcome even with the modern state-of-the-art
supercomputers.  The first problem is that they can only be implemented
on relatively small lattices and  at relatively high temperatures, since
the computational time increases rapidly with the number of sites of the
lattice, and as the temperature is lowered. Thus, ground state
properties in the thermodynamic limit can only be obtained via
finite-size scaling techniques and zero-temperature extrapolations.  The
second difficulty is the appearance of the sign problem when the system
is doped away from half-filling \cite{sign}.  The sign problem arises
from the fact that the fermion determinant, which is interpreted as a
probablity weight of the distribution of Hubbard-Stratonovich fields, is
not always positive definite, but decays exponentially with decreasing
temperature.  Thus, since averages must be computed by taking
differences of two positive quantities of similar magnitude, the
numerical problem becomes unstable, giving rise to large errors in the
measurements. The sign problem in the Hubbard model becomes worse at low
temperatures for values of the doping where a possible pairing phase is
predicted to exist.

The appearance of these problems has stimulated the search of new
techniques for solving Hubbard-like models beyond quantum Monte Carlo
simulations \cite{techniques,DMRG,Noack}.   In this paper we will
instead show that some of the above limitations can be partially
overcome using new types of boundary conditions (BCs) , {\it smooth
boundary conditions} (SBCs) , which have been successfully applied to the
study of one-dimensional (1D) systems within a large number of numerical
techniques \cite{SBCs} .  These BCs consist of {\it smoothly} decreasing
the energy parameters appearing in the Hamiltonian as we approach the
edges of the lattice.  The result of this operation is that, instead of
having a sharp and rigid boundary as is the case with open boundary
conditions (OBCs), with SBCs the boundary extends itself into the system
in such a way that its exact size is not fully determinable.  In
general, we will talk of the {\it bulk} of the system as the region
where the energy parameters are constant, and of the {\it boundary} as
the region over which the parameters are smoothly turned off. All
measurements are made in the bulk region. In addition to reaching the
thermodynamic limit on a relatively smaller system than with periodic
boundary conditions (PBCs)  or OBCs  we find that with SBCs we obtain an
improvement in the behavior of the sign problem which allows one to
reach significantly lower temperatures and to explore the possible
existence of a superconducting phase in the doped system.  In Section II
we introduce the Hubbard model with SBCs,  in Section III we demonstrate
the validity of SBCs comparing them to PBCs in the discussion of the
half-filled case, in Section IV we describe the behavior of the average
sign of the simulation with SBCs and study pairing correlations in the
nearly half-filled case, and, finally, in Section V we summarize and
conclude.  

\section{The model with smooth boundary conditions}

We consider the positive-$U$ Hubbard model \cite{Hubbard}
on a two-dimensional lattice defined by the Hamiltonian
\widetext
\begin{equation}
H=
-\sum_{\langle ij \rangle ,\sigma} t_{ij}(c^{\dagger}_{i\sigma}c_{j\sigma} +
c^{\dagger}_{j\sigma}c_{i\sigma})
-\sum_{i,\sigma} \mu_{i} n_{i\sigma}+
\sum_i U_{i} (n_{i\uparrow}-{1\over 2})(n_{i\downarrow}-{1\over 2}),
\label{H}
\end{equation}
\narrowtext
which consists of a system of itinerant electrons with an on-site
interaction of coupling strength $U_i$. Here $t_{ij}$ is the
nearest-neighbor hopping parameter between sites $i$ and $j$.  All the
above energy parameters appearing in the Hamiltonian are scaled
according to the smoothing function $f_i$, shown in Fig.\ \ref{fig1}, in
such a way that $U_i / U= \mu_i / \mu = f_i$ and $t_{ij} / t = {1 \over
2} (f_i + f_j)$, where $U$, $\mu$, and $t$ are the bulk values.  In the
following, without loss of generality, we will take $t=1$, and express
all the other parameters appearing in the Hamiltonian of Eq.(\ref{H}) in
dimensionless form.  The $c^\dagger_{i\sigma}$ are fermion creation
operators in a Wannier orbital centered at site $i$ with spin $\sigma$,
$\mu$ is the chemical potential, and
$n_{i\sigma}=c^\dagger_{i\sigma}c_{i\sigma}$ is the density operator.

The above rescaling of the energy parameters in the Hamiltonian is
motivated by a similar procedure that was successfully applied to
several 1D systems \cite{SBCs}. The choice of the smoothing function
$f_i$ emerges from a generalization of the 1D analogue, but is not
unique.   Here we use $f_i = y(1-d_i/[N_f+1])$, where $d_i$ measures the
distance of site $i$ from the nearest edge of the system, starting at 1
for $i$ on the outermost ``frame'', or square, of sites. $N_f$ is the
number of frames in the boundary region. The smoothing
function is given by\cite{SBCs}
\begin{equation}
y(x) = \left\{ \begin{array}{ll}
1&x<=0 \\
{1 \over 2} \left[ 1 + \tanh {x-1/2 \over x(1-x)} \right] & 0 < x < 1\\
0&x>=1.
\end{array} \right .
\label{ydef}
\end{equation}
The most important property of the the function $y(x)$ is that it and
all its derivatives are continuous everywhere. Thus in the limit of
a large number of frames there are no precise interfaces or boundaries
which would cause scattering. However,
for most of the results shown here, we use $N_f=2$, and $f_i$ takes on
the values 0.817574476 and 0.182425524. A surprising result of our
work is that one can obtain excellent smoothing using only two frames.

\section{The half-filled case}

It is well known that the half-filled Hubbard model on an infinite
lattice has an antiferromagnetic ground state with periodicity
commensurate with the lattice \cite{HubSC}.   
At finite temperature,
from symmetry considerations of the antiferromagnetic order parameter,
we know that the correlation length remains finite, and, thus, no phase
transition occurs.  On a finite lattice, however, the behavior is
different, since the correlation length can only grow as large as the
linear size of the system. Thus, we can still speak of an approximate
transition temperature for a finite lattice corresponding to where the
correlation length reaches the size of the lattice.  This different
behavior between a finite and infinite system is very crucial and could
lead to the erroneous conclusion that even the infinite system has a
finite transition temperature \cite{gapU}.   In general, finite-size
effects will be sensitive to the type of BCs used. Thus, it is important
to choose BCs which elminate finite-size effects already on relatively
small lattices so that the extrapolation to the infinite system can be
taken without the need to consider extremely large lattices.

Despite the fact that the behavior of the half-filled case is well
known, we are interested in studying it in the context of SBCs since we
want to show that we can obtain thermodynamic limit results on a
relatively smaller lattice than when using PBCs.  Also, since at
half-filling there is no sign problem, we want to show that this effect
is independent of the behavior of the average sign with SBCs relative to
PBCs.   We find that applying SBCs to a finite lattice indeed reduces
finite-size effects and allows faster convergence to the thermodynamic
limit.  We consider two types of measurements, local quantities and
correlation functions.  We first consider the average kinetic energy per
site, given by
\widetext
\begin{equation}
\langle K_i \rangle = -{1\over 4}\langle \sum_{j,\sigma} t_{ij} 
(c_{i_\sigma}^\dagger c_{j\sigma} + c_{j_\sigma}^\dagger c_{i\sigma} 
)\rangle,
\label{kinetic}
\end{equation}
\narrowtext
where the sum runs over all $j$ nearest neighbors of $i$.  In Fig.\
\ref{fig2} we show $\langle K_i \rangle$ as a function of the linear
size of the system using PBCs and SBCs.  The results for SBCs are obtained
with the number of smoothing frames $N_f$ fixed at 2.
This means that the bulk will have a linear size
that is 4 sites smaller than the corresponding system with PBCs. It is
very striking that already on a $6\times 6$ system with a $2\times 2$
bulk region SBCs give a relatively good estimate to the value that we
obtain on a $16\times 16$ lattice with PBCs.  A similar conclusion can be
drawn from Fig.\ \ref{fig3}, where we show the double occupancy $\langle
n_{i\uparrow} n_{i\downarrow} \rangle$.  The finite-size effects in this
case are much stronger with PBCs than SBCs. Notice that both in Figs.\
\ref{fig2} and \ \ref{fig3} the measured quantities saturate to the same
value for either type of BCs as the system size becomes larger, as
expected, since the type of BCs should not have much effect on the
behavior of a very large system. 

We also consider several other types of local measurements, leading to
the same conclusion: {\it the thermodynamic limit is reached on a
smaller lattice with SBCs rather than with PBCs or OBCs}.  In Fig.\
\ref{fig4} we show the average total energy as a function of the inverse
temperature. Also in this case, we see that for each temperature
considered, with SBCs we reach the thermodynamic limit more rapidly than
with PBCs.  This behavior is very similar to what we found in the
numerical density-matrix renormalization group (DMRG) study of the
spin-$1 \over 2$ Heisenberg chain with SBCs.  Also in that case, with SBCs
the ground-state energy is reached on smaller lattices than with OBCs
\cite{SBC2}.

We also examine various correlation functions in order to show that the
behavior is similar to that of the local quantities.  When using SBCs the
correlation functions can only be evaluated for two sites separated by a
distance only as large as the size of the bulk.  Similarly, with PBCs we
can only calculate correlations between points as far out as half the
linear size of the system, since the lattice wraps around onto itself.
An additional difference is that, since SBCs break the translational
invariance of the Hamiltonian, the measurement of local quantities and
correlation functions cannot be averaged over the entire lattice, as can
be done with PBCs.   In this case, the averaging can only be done over
sites, or pairs of sites, which are equivalent under the reduced
symmetry of the lattice (reflection and inversion) with SBCs.  This
implies that, in order to obtain better statistics, we need to run the
simulation longer than with PBCs. This is probably the only significant
disadvantage of SBCs over PBCs.  In Fig.\ \ref{fig5} we show the spin-spin
correlation function, defined by 
\widetext
\begin{equation}
c(l)=c(l_x,l_y)=\langle (n_{i+l,\uparrow}-n_{i+l,\downarrow})
(n_{i,\uparrow}-n_{i,\downarrow}) \rangle.
\label{spincorr}
\end{equation}
\narrowtext
for several systems sizes with PBCs and compare it to the case with SBCs
on a $10\times 10$ lattice with two smoothing frames.   Again, it
appears that with SBCs we can obtain on a smaller system the same results
as with PBCs on a considerably larger system.   Having shown that the use
of SBCs allows one to reduce finite-size effects on relatively small
lattices, we now study the effect of SBCs on the average sign by
considering the Hubbard model away from half-filling.
 
\section{The doped system}

As one dopes the system, the long-range antiferromagnetic order which is
present in the ground state of the half-filled band is abruptly
destroyed and short-range incommensurate magnetic correlations set in
\cite{Schulz}.   Here we are mainly interested in studying the possible
existence of a superconducting phase away from half-filling.  Recently,
there has been an enormous effort in trying to obtain the phase diagram
of the doped Hubbard model. Most of the studies have not been able to
determine conclusively whether there is or there is not a
superconducting phase.  Approximate techniques have been used, such as
self-consistent calculations, mean-field theories and conserving
approximations \cite{techniques}, often suggesting the possibility of a
parameter regime having $d$-wave superconductivity.  QMC results, which
show evidence of an attractive pairing interaction, but no direct
evidence of pairing, have not been completely satisfactory because of
the appearance of the sign problem already at relatively high
temperatures. The possibility exists that the temperature at which the
pairing occurs is significantly lower than the accessible temperatures
even on as small lattices as an $8\times 8$.   

Although initially we did not expect changing boundary conditions would
have a large effect on the average sign, in fact it can.  In general,
the average sign decays exponentially with the inverse temperature
$\beta$ and system size $N$, $\langle S \rangle \sim e^{-A\beta N}$
\cite{sign}, where $A$ depends on the QMC method (determinantal,
world-line, projector, {\it etc.}).  If $\langle S \rangle$ falls below
about 0.1, it becomes impossible to obtain reasonable results.
Previously, we showed that for large values of $U$ the sign decays
slower in the world-line method than in the determinantal approach
\cite{worldmc}.  Here we find that the decay of the average sign can be
substantially slower with SBCs than with either open BCs or with PBCs.

In Fig.\ \ref{fig6} we show the average sign as a function of inverse
temperature for various lattice sizes at a filling of $\langle n \rangle
= 0.87$ with $U=4$ for several types of BCs. If we consider the case with
OBCs and PBCs only, we see that on an $8\times 8$ lattice the average sign
is already too small at a $\beta \approx 6-7$ for calculating
measurements accurately. On the other hand, when we consider SBCs we see
that the sign allows one to go to lower temperatures even on a
lattice as large as a $12\times 12$ with two smoothing frames. In
general we find that {\it the average sign as a function of temperature
decays slower with SBCs than with PBCs and OBCs for all values of U}.
Particularly suprising is that the smooth system with a {\it bulk}
region of size $4\times4$ (the solid triangles) has a sign that decays
substantially more slowly than either a periodic or open system whose {\it
full}  size is $4\times4$.

There are several ways in which one can understand why the average sign
is better behaved with SBCs rather than with other types of BCs.
When we decrease the energy parameters in the Hamiltonian on the edges of
the system, we effectively make the local temperature there higher. Thus,
smoothing the energy parameters of
the system is equivalent to introducing a temperature gradient on
the lattice without any heat flow into or out of the bulk of the system.
(This strange temperature gradient cannot be realized in a real system,
as far as we know.)
Thus, if one thought of the average sign as a measurement that is
obtained from averaging over the entire lattice, it might be
reasonable to expect its behavior to be better
with SBCs rather than PBCs. However, the average sign does not really
involve some averaging process over the entire lattice---something slightly
more subtle is going on.

However, this point of view can give us some insight into understanding why
finite-size effects are much smaller with SBCs rather than with PBCs.
Since the boundary is at a relatively higher temperature, a larger number
of local states in the Hilbert space are accessible there. Thus, on the
boundary, the system has the freedom to choose a larger portion of the
Hilbert space, and thus, it can adjust itself in order to satisfy the
constraints of the bulk. In the partition function, the lower temperature
of the bulk region dominates the high temperature edge region, so the
edges adjust to satisfy the bulk.

As mentioned above, the most surprising effect that emerges from 
Fig.\ \ref{fig6} is that the
sign on an $8\times 8$ system with SBCs and a $4\times 4$ bulk region is
better behaved than on a $4\times 4$ system with OBCs or PBCs. At this point
we only have a qualitative argument for this effect.
Based on the projector QMC simulation of the same system,
we can at least show that this behavior is not completely unexpected.
We have verified that the average sign in the projector QMC
has a very similar behavior, namely that it is improved with SBCs.
In the projector QMC the sign problem
originates from the phase of matrix
elements (entering in the partition function) of the form,
\widetext
\begin{equation}
\langle \phi(0) | \phi(\tau)\rangle \equiv
\langle \phi(0) | e^{-\tau  H} | \phi(0) \rangle,
\label{projector}
\end{equation}
\narrowtext
where
$|\phi(0) \rangle$ is an initial trial state
(which can be, for example, a filled Fermi sea), and $|\phi(\tau) \rangle$ 
is the
initial state propagated to imaginary time $\tau=L\Delta \tau$.
The fluctuating Hubbard-Stratonovich fields cause
the state $| \phi(\tau) \rangle$ to evolve through 
Hilbert space, and its precise evolution will depend on the type of BCs used.
In particular, with SBCs,
in the boundary region, where the temperature is high, the
system evolves slowly, so that the propagated state is still very close
to the initial state there.
The state could evolve rapidly in the bulk region, except for the fact
that it is continuously connected to the boundary region. 
Thus the edges act as
a drag force on the bulk, slowing the rapid variations in imaginary
time that cause the sign problem.

In order to verify this picture, we studied, using the projector method,
a simplified ``toy'' model: a 2D non-interacting tight-binding
system with the addition of a Ising-like field $\Delta_i=\pm 1$
coupled to the
density operator $n_i$, designed to mimic the Hubbard-Stratonovich field:
\widetext
\begin{equation}
H=-\sum_{\langle i,j\rangle} t_{ij}
\left( c_i^\dagger c_j+c_j^\dagger c_i\right)
+\sum_i\Delta_in_i.
\label{TBH}
\end{equation}
\narrowtext
Two particles were put in the system, since a single particle never
has a sign problem. The state of the higher energy particle had
a nodal line which moved as the system evolved through imaginary time.
If the nodal line rotated through 180$^\circ$, the system had 
a minus sign. A specific field configuration was chosen with a spiral
configuration through imaginary time to try to drag the particles 
in a circle in order to get a minus sign as quickly as possible.
The field was able to cause rapid rotation of the nodal line in
both PBCs and OBCs. In SBCs, however, the nodal line was unable
to evolve rapidly on the edges (it was stuck, as if in a viscous fluid), 
and the part of the
nodal line in the bulk was held back. Consequently, it was much 
more difficult to generate configurations corresponding to minus signs.
We believe the results support our picture for the improved behavior
of the sign in the interacting system.

We now present local measurements and correlation functions at high
enough temperatures to be accessible to both PBCs and SBCs. 
When comparing the results with the two types of BCs we find 
similar behavior to the half-filled case. Then we will present results
at lower temperatures, unaccessible with PBCs, and continue the analysis
with SBCs. 

In Fig.\ \ref{fig7} we show the kinetic energy as a function of inverse
temperature at a filling of $\langle n \rangle =0.87$, with $U=4$. 
With PBCs we can reach
only a temperature of $\beta = 6$ on an $8\times 8$ lattice, while with
SBCs we can reach $\beta = 12$.

We considered three types of pairing correlation functions, corresponding to
different symmetries of the order parameter. In general, a given 
pairing correlation function is given by 
\widetext
\begin{equation}
D(l)=|\langle \Delta_{i+l}\Delta_i^\dagger \rangle|,
\label{paircorr}
\end{equation}
\narrowtext
where the pairing field operators, $\Delta_l$, are given by
\widetext
\begin{equation}
\Delta^s_l=c_{l\uparrow}c_{l\downarrow},
\label{spairing}
\end{equation}
\narrowtext
for $s$-wave symmetry, and by
\widetext
\begin{equation}
\Delta^{s^*}_l= 
{1\over 2}\left( 
c_{l\uparrow}c_{l+x\downarrow}+c_{l\uparrow}c_{l-x\downarrow}\right) +
{1\over 2}\left(
c_{l\uparrow}c_{l+y\downarrow}+c_{l\uparrow}c_{l-y\downarrow}\right),
\label{sextpairing}
\end{equation}
\narrowtext
and
\widetext
\begin{equation}
\Delta^{d}_l= 
{1\over 2}\left(
c_{l\uparrow}c_{l+x\downarrow}+c_{l\uparrow}c_{l-x\downarrow}\right) -
{1\over 2}\left(
c_{l\uparrow}c_{l+y\downarrow}+c_{l\uparrow}c_{l-y\downarrow}\right),
\label{dpairing}
\end{equation}
\narrowtext
for extended $s$-wave and $d$-wave symmetry, respectively.

In Fig.\ \ref{fig8} we show the pairing correlation functions for the above 
three different symmetry channels. For all temperatures considered, we find that
pairing in a $d$-wave channel is always an order of magnitude stronger than 
for an $s$-wave or extended $s$-wave channels. 
Thus, we
continue our analysis with the $d$-wave pairing correlation function only,
examine its temperature dependence and compare it to the non interacting
case to see whether there is an enhancement with respect to the $U=0$ case. 
In Fig.\ \ref{fig9}
we show the $d$-wave pairing correlation function for
several temperatures with $U=8$, 
showing that, as the temperature is decreased there is an
enhancement in the pairing. For reference, we also show the corresponding $U=0$
results to show that there is no enhancement relative to the non-interacting
case, for the accessible temperatures.
In Fig.\ \ref{fig10}
we show the $d$-wave pairing correlation function for
lower temperatures with $U=4$. Here, at $\beta=10$, we do find evidence
for enhancement over the non-interacting case, near the M point.
This may be a sign of $d$-wave superconductivity.

\section{Conclusions}

We have implemented a numerical simulation of the 2D Hubbard
model using SBCs. We have showed that at half-filling,
where there is no sign problem, one obtains thermodynamic limit
results on a smaller lattice than when using PBCs. Away from half-filling,
we have found that the average sign decays more slowly with inverse temperature
and lattice size with SBCs than with OBCs and PBCs, allowing us to reach
significantly lower temperatures and larger lattices. We looked at the
pairing correlation functions and showed that the $d$-wave channel is
favored over other types of pairing channels, and that the pairing
increases as we lower the temperature. 
On a $10\times 10$ lattice 
for $U=4$, at a inverse temperature of
$\beta=10$, we did find enhancement of the $d$-wave
correlations with respect to the non-interacting case, a possible
sign of $d$-wave superconductivity.

\section{Acknowledgements}

The authors would like to thank R.M.\ Noack and R.T.\ Scalettar for 
useful discussions.
We acknowledge support from the Office of Naval
Research under grant No. N00014-91-J-1143.
The numerical calculations reported in this paper were performed at
the San Diego Supercomputer Center.

\begin{figure}
\caption{
The smoothing function, $f_i=f_{i_x,i_y}$ for a $12\times 12$ lattice with
four smoothing frames. The bulk region is a $4\times 4$ lattice.  The four
concentric square frames are clearly visible.
}
\label{fig1}
\end{figure}

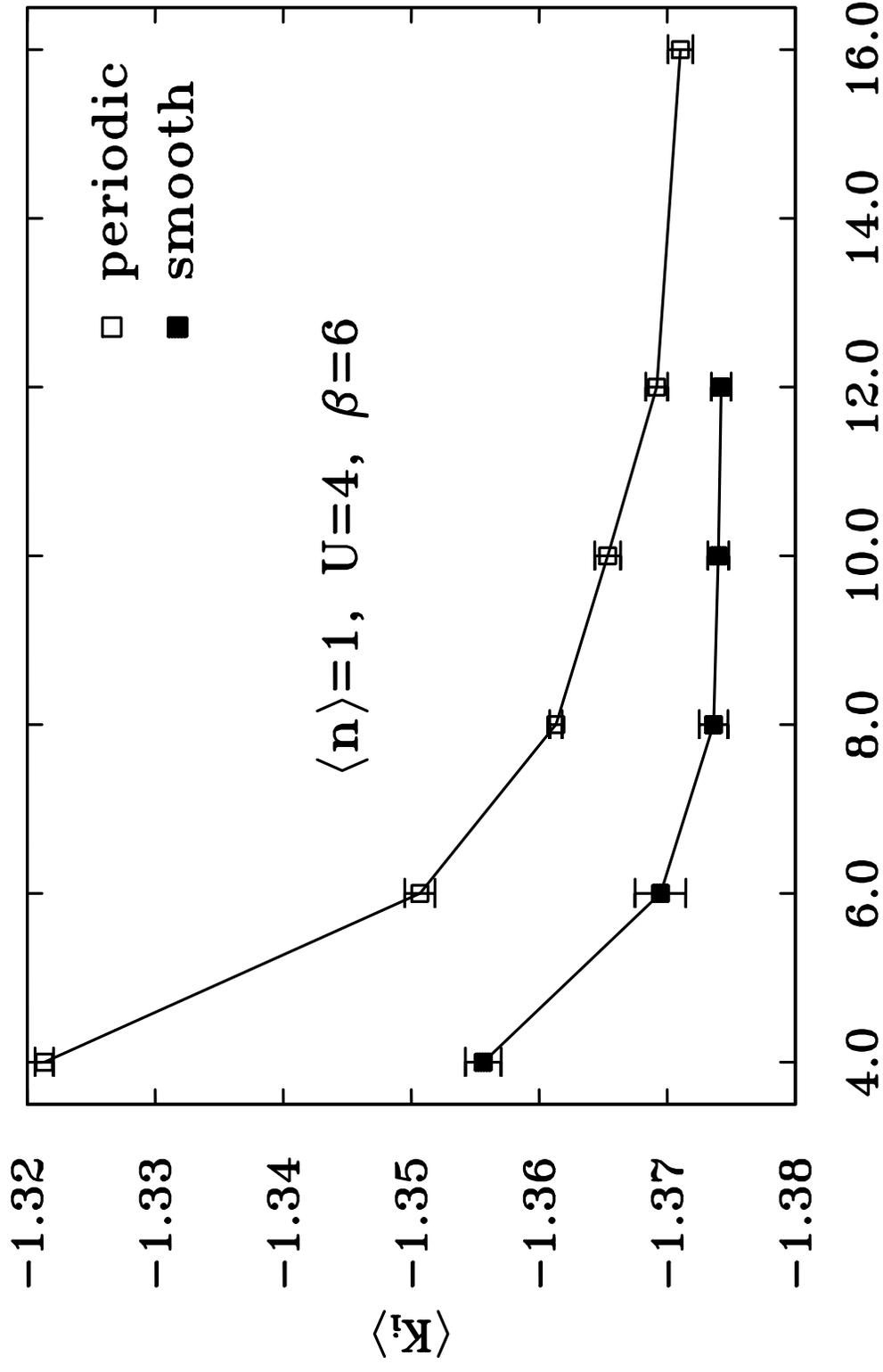
\begin{figure}
\caption{
The
kinetic energy, $\langle K_i \rangle$, measured at the center of the 
lattice, as a function of linear system size 
$N_x$ with $U=4$ at half-filling at $\beta=6$. The solid squares are with
PBCs and the solid circles with SBCs with two smoothing frames.
}
\label{fig2}
\end{figure}

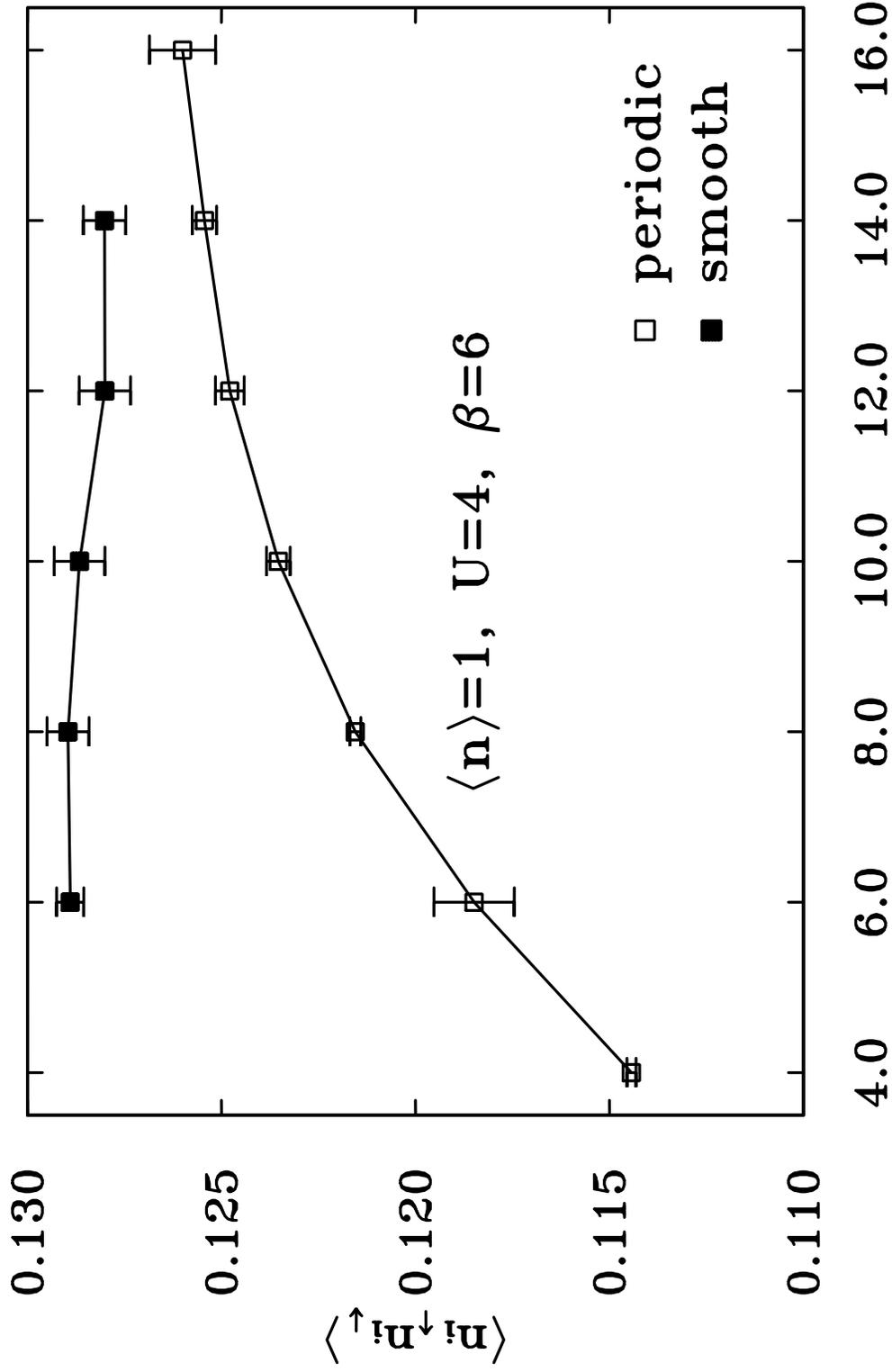
\begin{figure}
\caption{
The double occupancy $\langle n_{i\uparrow}n_{i\downarrow}\rangle$,
measuread at the center of the lattice, as a 
function of linear system size $N_x$ with $U=4$ at half-filling at $\beta=6$.
The solid squares are with PBCs and the solid circles with SBCs with two
smoothing frames.
}
\label{fig3}
\end{figure}

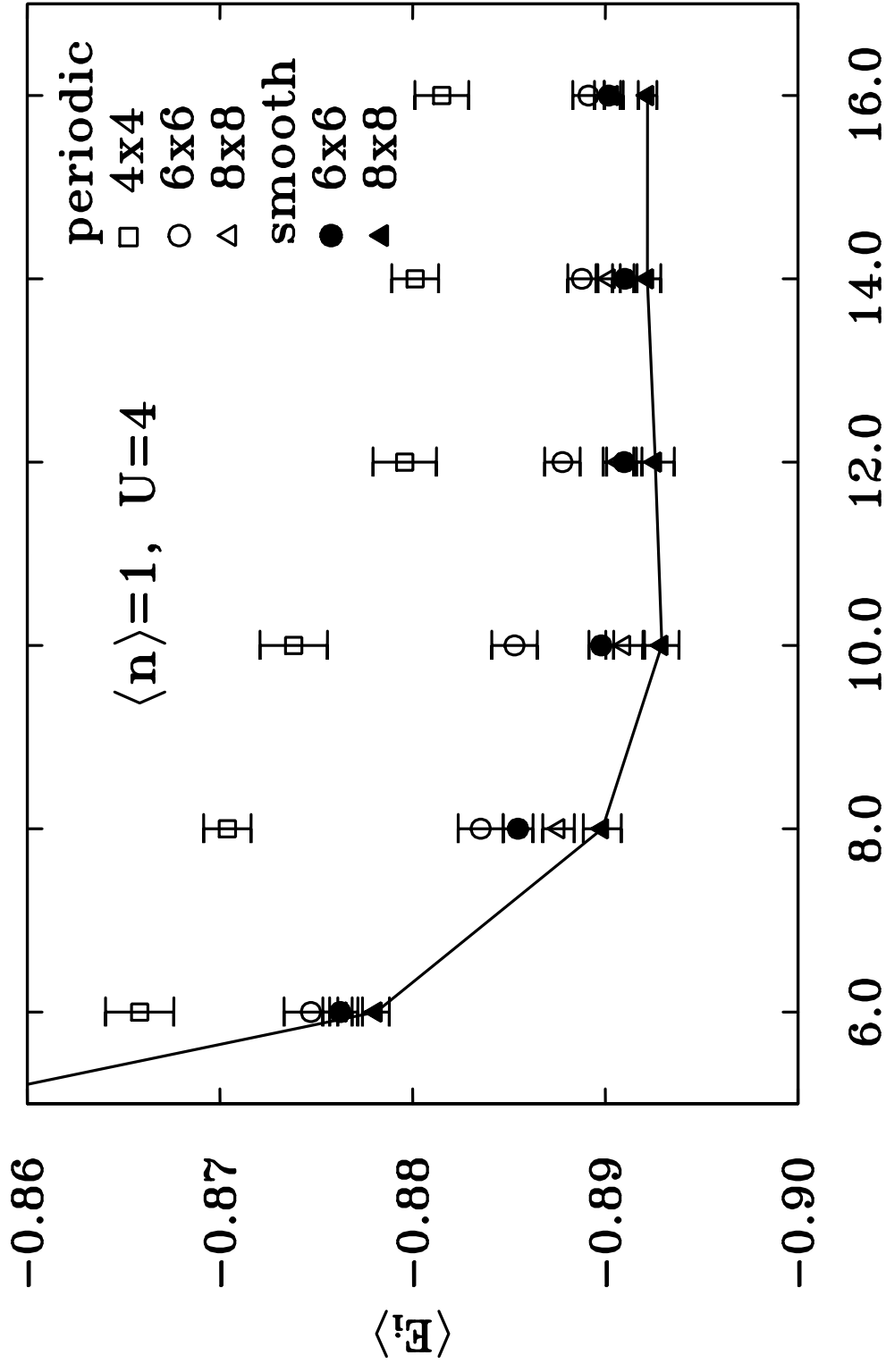
\begin{figure}
\caption{
The total energy $\langle E_i \rangle$, measuread at a site $i$ in the 
center of the lattice, as a function of inverse temperature 
$\beta$ for several system sizes and with PBCs and SBCs. The filling is
$\langle n\rangle = 1$ and $U=4$. The open symbols are with PBCs and the solid 
symbols with SBCs.   With SBCs we use two smoothing frames on the boundary.
The solid lines are just guides to the eye.
}
\label{fig4}
\end{figure}

\begin{figure}
\caption{
The spin-spin correlation function $c(l_x,l_y)$ at half-filling with $U=4$
and at $\beta=6$. The path taken in calculating $c(l_x,l_y)$ is shown in the
inset. The solid squares are for a 
$10\times 10$ system with SBCs and two smoothing frames. The open symbols
are with PBCs and on sizes as shown. The solid line is a guide to the eye for
the case with SBCs. The errors (not shown) are smaller than the size of the 
symbols.
}
\label{fig5}
\end{figure}

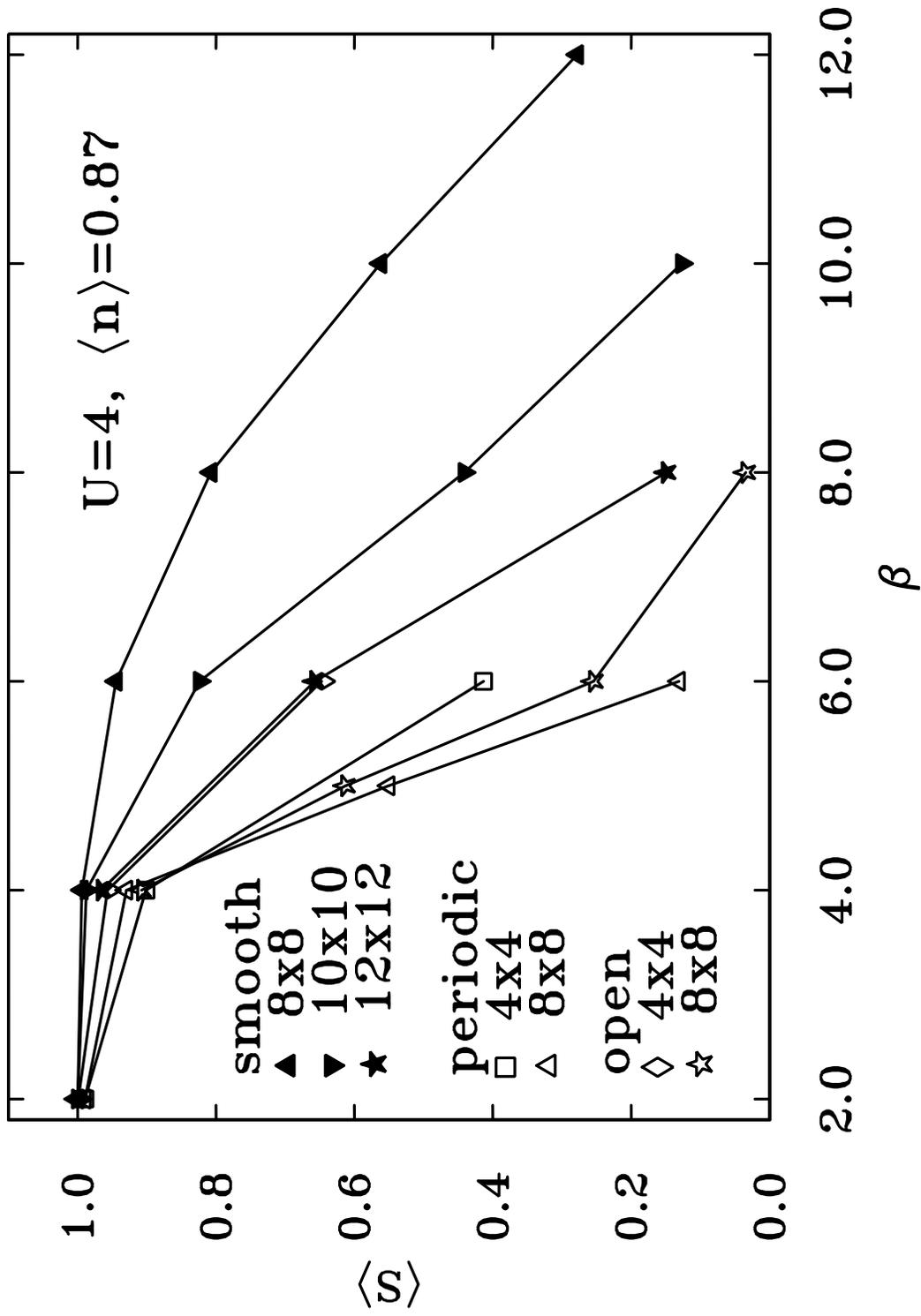
\begin{figure}
\caption{
The average sign $\langle S \rangle$ as a function of $\beta$ with
$U=4$ and at $\langle n \rangle = 0.87$ for different system sizes and 
BCs. The open symbols are for OBCs and PBCs and the
solid symbols are for SBCs for the sizes indicated. The solid lines are just
guides to the eye.
}
\label{fig6}
\end{figure}

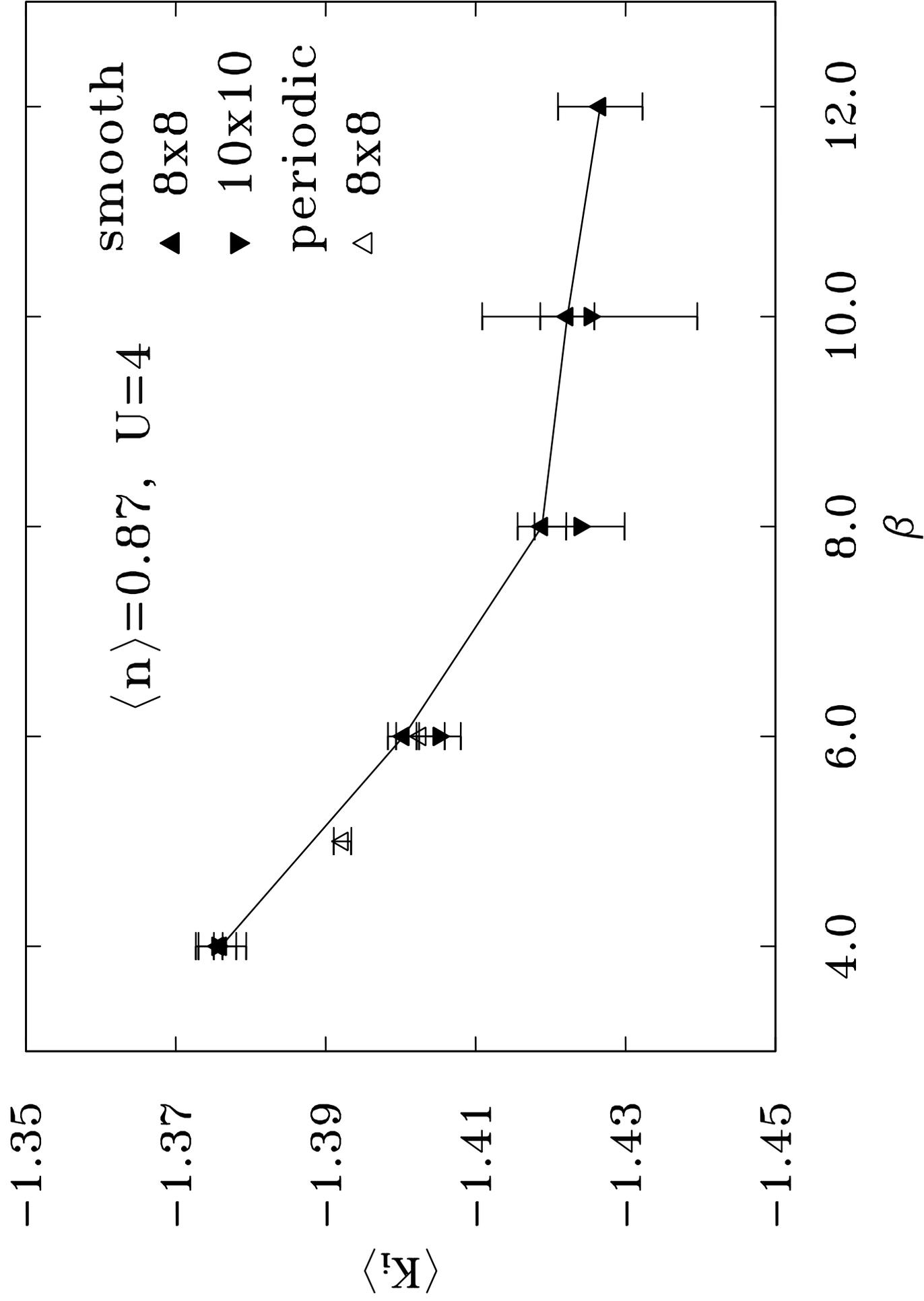
\begin{figure}
\caption{
The kinetic energy $\langle K_i \rangle$, measuread at the center of the 
lattice, as a function of $\beta$ for
several systems sizes and with PBCs and SBCs.
The filling is $\langle n \rangle = 0.87$ and $U=4$. 
The open symbols are with PBCs and the solid
symbols with SBCs. The triangles are for an $8\times 8$ and the
upside-down triangles for a
$10\times 10$ lattices. With SBCs we use two
smoothing frames on the boundary.
}
\label{fig7}
\end{figure}

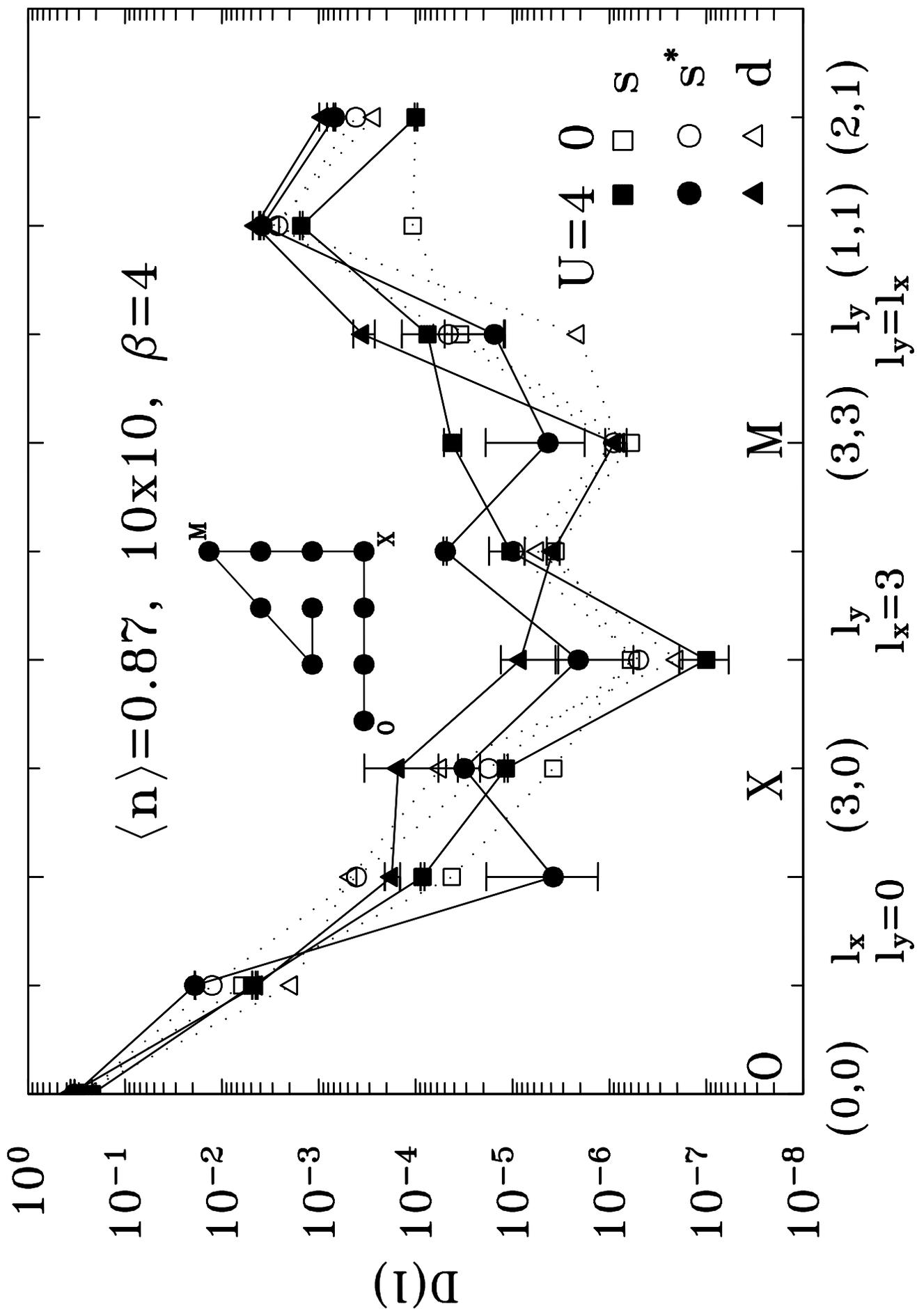
\begin{figure}
\caption{
The pairing correlation functions $D(l)$
with $U=4$ at $\langle n \rangle = 0.87$
and at $\beta=8$. The solid circles are for $s$-wave, the solid triangles are
for extended $s$-wave, and the solid squares are for $d$-wave. The path
taken through the lattice corresponds to a triangle as indicated in the inset.
The corresponding empty symbols are for $U=0$.
The solid lines are just guides to the eye.
}
\label{fig8}
\end{figure}

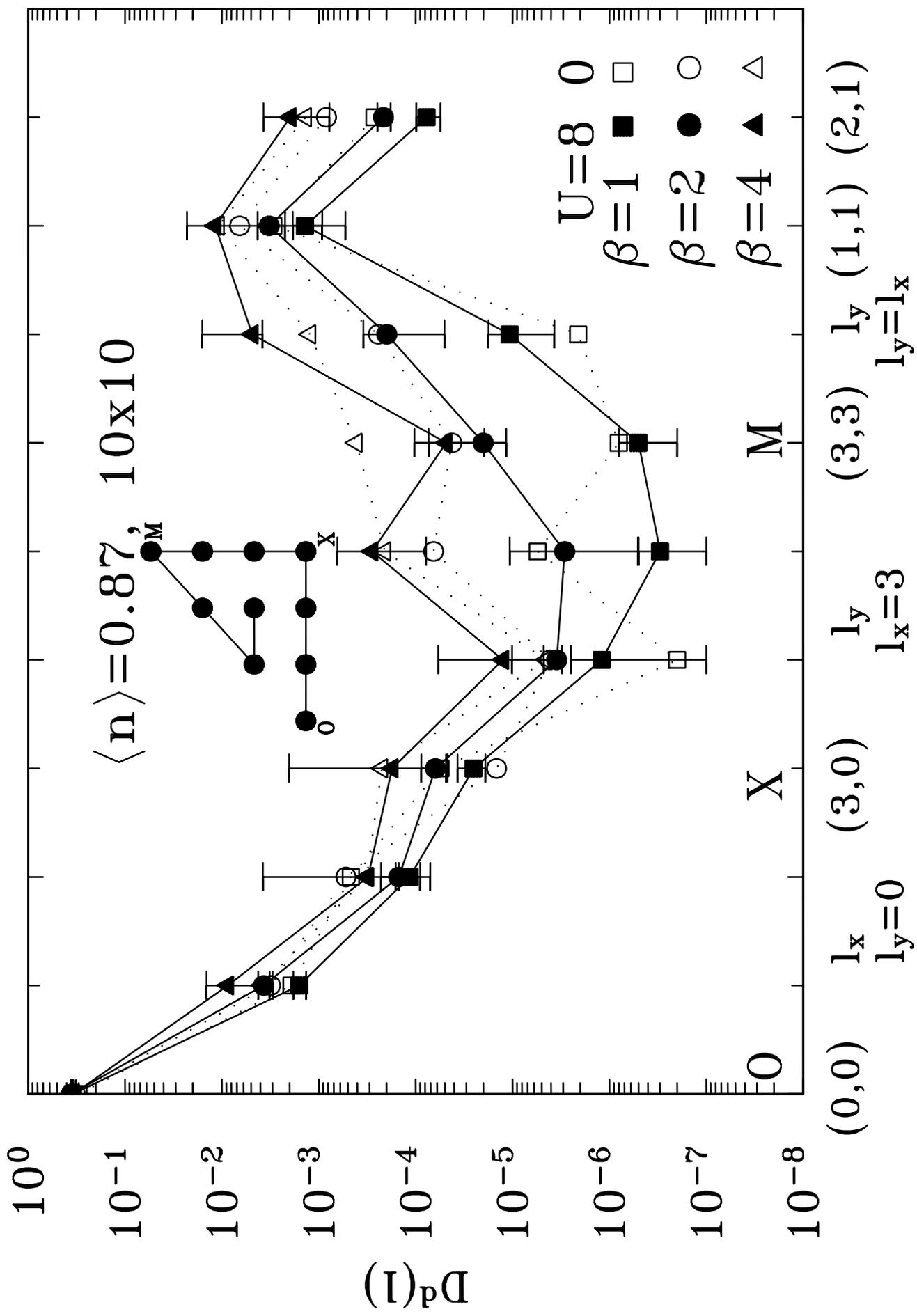
\begin{figure}
\caption{
The $d$-wave pairing correlation function $D^d(l)$
on a $10\times 10$ lattice with
$U=8$ at $\langle n \rangle = 0.87$ for several values of $\beta$. The path
taken through the lattice corresponds to a triangle as indicated in the inset.
The corresponding empty symbols are for $U=0$.
}
\label{fig9}
\end{figure}

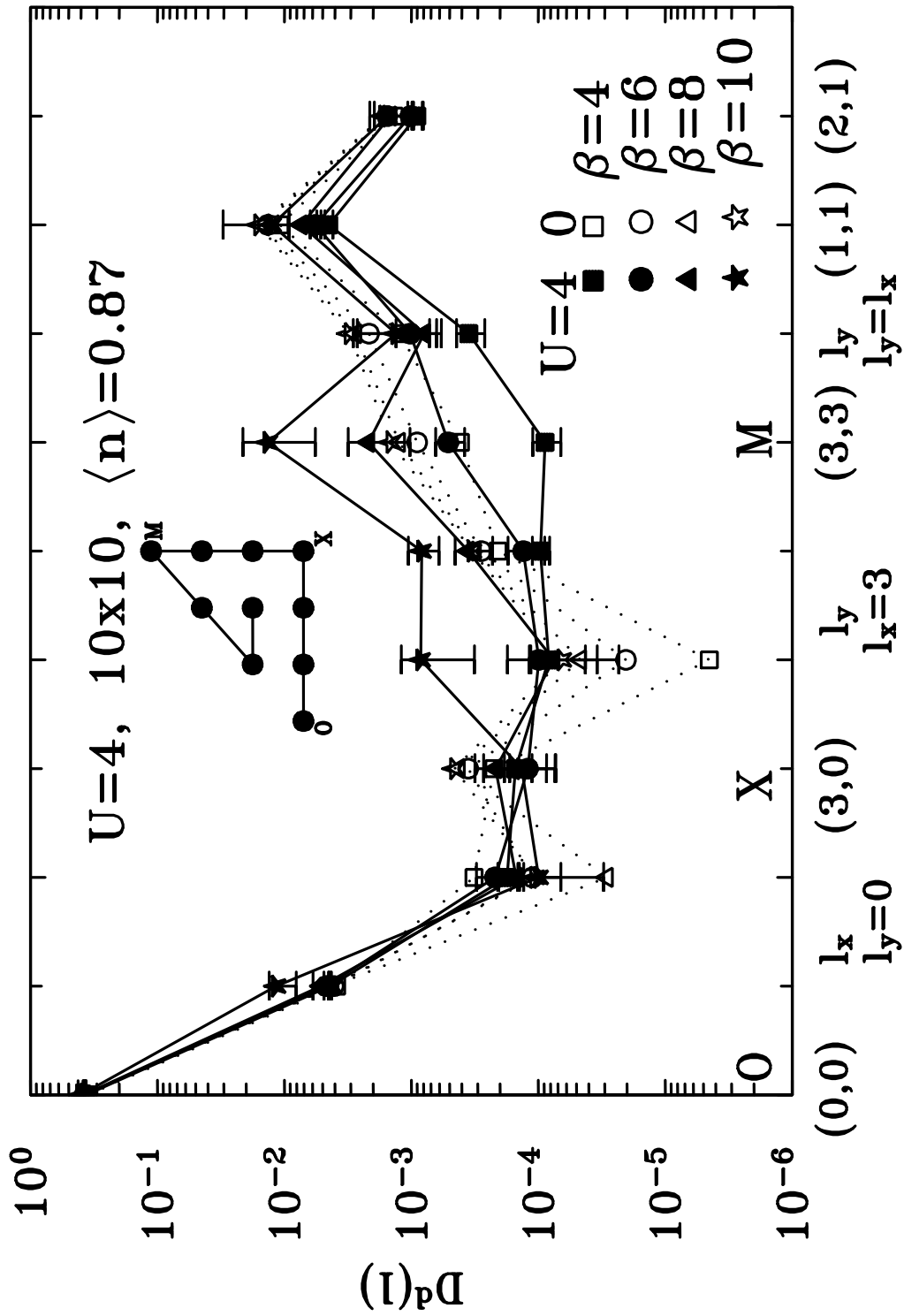
\begin{figure}
\caption{
The $d$-wave pairing correlation function $D^d(l)$
on a $10\times 10$ lattice with
$U=4$ at $\langle n \rangle = 0.87$ for several values of $\beta$. The path
taken through the lattice corresponds to a triangle as indicated in the inset.
The solid lines are just guides to the eye. The solid squares are for
$\beta=4$, the solid circles are for $\beta = 6$, the solid triangles are
for $\beta = 8$ and the open squares are for $\beta = 10$.
The corresponding empty symbols are for $U=0$.
}
\label{fig10}
\end{figure}

\end{document}